\documentclass[twocolumn,english]{revtex4}
\usepackage[T1]{fontenc}
\usepackage[latin9]{inputenc}
\usepackage[a4paper]{geometry}
\usepackage{bm}
\usepackage{amsmath}
\usepackage{amssymb}
\usepackage{graphicx}
\usepackage{esint}

\makeatletter
\@ifundefined{textcolor}{}
{%
 \definecolor{BLACK}{gray}{0}
 \definecolor{WHITE}{gray}{1}
 \definecolor{RED}{rgb}{1,0,0}
 \definecolor{GREEN}{rgb}{0,1,0}
 \definecolor{BLUE}{rgb}{0,0,1}
 \definecolor{CYAN}{cmyk}{1,0,0,0}
 \definecolor{MAGENTA}{cmyk}{0,1,0,0}
 \definecolor{YELLOW}{cmyk}{0,0,1,0}
}

\makeatother

\usepackage{babel}
\begin{document}

\title{Scattering of universal fermionic clusters in the resonating group
method}

\author{Pascal Naidon$^{1}$, Shimpei Endo$^{2}$, and Antonio M. Garc{\'i}a-Garc{\'i}a$^{3}$}

\affiliation{$^{1}$RIKEN Nishina Centre, RIKEN, Wako, 351-0198 Japan}

\affiliation{$^{2}$Laboratoire Kastler-Brossel, {\'E}cole Normale Sup{\'e}rieure, 24
rue Lhomond, 75231 Paris, France}

\affiliation{$^{3}$University of Cambridge, Cavendish Laboratory, JJ Thomson
Avenue, Cambridge, CB3 0HE, UK}

\date{\today}
\begin{abstract}
Mixtures of polarised fermions of two diff{}erent masses can form
weakly-bound clusters, such as dimers and trimers, that are universally
described by the scattering length between the heavy and light fermions.
We use the resonating group method to investigate the low-energy scattering
processes involving dimers or trimers. The method reproduces approximately
the known particle-dimer and dimer-dimer scattering lengths. We use
it to estimate the trimer-trimer scattering length, which is presently
unknown, and find it to be positive. 
\end{abstract}
\maketitle

\section{Introduction}

In the last decade, the use of controlled Feshbach resonances in ultra-cold
atom experiments have enabled the study of low-energy quantum systems
of particles interacting with large scattering lengths. Close to a
Feshbach resonance, the interparticle scattering length is much larger
than the range of interparticle forces. As a result, the low-energy
properties of these systems are universal, in the sense that they
depend only upon a few parameters, such as the scattering length \cite{Braaten2006},
and the three-body parameter \cite{Wang2012,Naidon2014a} in systems
exhibiting the Efimov effect~\cite{Efimov1970a,Efimov1970b,Efimov1973}.
Moreover, close to Feshbach resonances, atoms can be associated into
clusters of universal character: diatomic molecules called Feshbach
molecules that are a realisation of universal dimers \cite{Herbig2003,Regal2003,Duerr2004,Cubizolles2003,Jochim2003,Strecker2003,Zwierlein2003,Ospelkaus2006},
triatomic molecules that are a realisation of Efimov states \cite{Kraemer2006,Lompe2010a,Nakajima2011}.
Theory predicts the existence of a variety of other universal clusters
of larger number of particles \cite{Stecher2009,Stecher2010,Blume2012}
that are expected to be observed experimentally in the future~\cite{Ferlaino2009}.

The few-body properties, in particular the scattering properties of
clusters, can play a crucial role in the identification and stability
of the many-body ground states of these systems. For instance, the
stability of a gas of universal dimers made of fermions was observed
\cite{Cubizolles2003,Strecker2003,Jochim2003,Regal2003,Zwierlein2003,Ospelkaus2006}
and explained theoretically \cite{Petrov2004a,Marcelis2008} by exact
four-body calculations for two scattering dimers. 

Although it is sometimes feasible to calculate exactly the wave function
of an $N$-body cluster \cite{Efimov1970a,Kartavtsev2007}, the exact
computation of the scattering properties of two clusters is generally
out of reach for $N\ge3$.\textbf{ }In the context of nuclear and
sub-nuclear physics, a broad array of approximation schemes have been
successfully developed to address similar problems. One of the leading
techniques is the so-called Resonating Group Method (RGM), introduced
by Wheeler~\cite{Wheeler1937}, to study light nuclei, such as $^{16}\mbox{O}$
and $^{8}\ensuremath{\mbox{Be}}$, modelled as clusters of $\alpha$
particles. Since then, it has been employed in a variety problems
including the scattering of light nuclei, the stability of light nuclei
to external nucleon scattering and nuclear particles \cite{Thompson1977,Tang1978}.
More recently, it has been used \cite{Faessler1984,Shimizu1989,Oka2000}
to study low-energy scattering, and bound states, of baryon-baryon
and other multi-quark cluster configurations.

In the single-channel approximation, the RGM constructs the low-energy
scattering wave function of two or more scattering clusters from the
wave functions of the individual clusters, while preserving the full
antisymmetrization of wave functions. This gives an effective potential
between the clusters that can be used to treat scattering as well
as bound states. It is especially accurate  in situations in which
single clusters are not strongly altered by the scattering process.
Here, we propose to apply this method to the low-energy scattering
of universal fermionic clusters that are relevant to ultra-cold atoms
close to Feshbach resonances.

The paper is organised as follows. In section~\ref{sec:The-Resonating-Group},
we review the essence of the RGM. In section~\ref{sec:Scattering-of-dimers},
we apply it to universal clusters whose scattering properties are
known. In section~\ref{sec:Scattering-of-trimers}, we apply the
RGM to the yet unknown scattering of universal trimers.

\section{The Resonating Group Method\label{sec:The-Resonating-Group}}

\subsection{General formalism}

Let us consider the scattering between a cluster A of $n$ particles
and a cluster B of $N-n$ particles. It is assumed that the wave functions
$\phi_{\tiny\mbox{A}}(1,2,\dots,n)$ and $\phi_{\mbox{\tiny B}}(n+1,n+2,\dots,N)$
of these clusters are known. In the single-channel RGM, the $N$-body
wavefunction $\Psi$ describing the scattering process is constructed
as the antisymmetrised product of the cluster wave functions and a
wave function $\psi(\bm{R})$ for the relative motion between the
two clusters: 
\begin{equation}
\Psi=\mathcal{S}\left[\phi_{\mbox{\tiny A}}(1,\,2,\dots,\, n)\phi_{\tiny\mbox{B}}(n+1,\, n+2,\dots,\, N)\psi(\bm{R})\right].\label{eq:TotalWavefunction}
\end{equation}
Here, $\mathcal{S}$ denotes the symmetrisation (or antisymmetrisation)
operator that symmetrises (or antisymmetrises) the wave function under
the exchange of identical particles. Symmetrisation is performed for
bosonic particles, whereas antisymmetrisation is performed for fermionic
particles. The vector $\bm{R}$ describes the relative position between
the centres of mass of the two clusters. The idea behind this approximation
is that the structure of the two clusters is not much altered during
the collision, and the two clusters mix only through the exchange
of identical particles.

The purpose of the RGM is to determine the wave function $\psi(\bm{R})$
for the relative motion of the clusters. This is done by applying
the variation principle to the average quantity,
\begin{equation}
\langle\Psi\vert H-\mathcal{E}\vert\Psi\rangle,\label{eq:VariationalPrinciple}
\end{equation}
where $H$ is the total hamiltonian and $\mathcal{E}$ is the total
energy of the system. Requiring $\psi$ to extremise the above quantity
implies that for an infinitesimal variation $\delta\psi$ around $\psi$
we have
\[
\langle\mathcal{S}[\phi_{\mbox{\tiny A}}\phi_{\mbox{\tiny B}}\delta\psi]\vert H-\mathcal{E}\vert\mathcal{S}[\phi_{\tiny\mbox{A}}\phi_{\mbox{\tiny B}}\psi]\rangle+\mbox{c.c.}=0.
\]
The variations $\delta\psi$ and its complex conjugate $\delta\psi^{*}$
can be formally taken as independent variations, resulting in the
following Euler-Lagrange equation of motion, 
\begin{equation}
\langle\mathcal{S}[\phi_{\mbox{\tiny A}}\phi_{\mbox{\tiny B}}]\vert H-\mathcal{E}\vert\mathcal{S}[\phi_{\mbox{\tiny A}}\phi_{\mbox{\tiny B}}\psi]\rangle=0,\label{eq:EquationOfMotion0}
\end{equation}
which can be simplified as
\begin{equation}
\langle\phi_{\mbox{\tiny A}}\phi_{\mbox{\tiny B}}\vert H-\mathcal{E}\vert\mathcal{S}[\phi_{\mbox{\tiny A}}\phi_{\mbox{\tiny B}}\psi]\rangle=0.\label{eq:EquationOfMotion}
\end{equation}
since the total hamiltonian $H$ is invariant under the exchange of
identical particles. 

The hamiltonian $H$ consists of kinetic operators $t_{i}$ for each
particle and pairwise interaction terms $V_{ij}$ for each pair of
particles,
\begin{equation}
H=\sum_{i=1}^{N}t_{i}+\sum_{i<j}^{N}V_{ij}-t_{\mbox{\tiny c}}.\label{eq:Hamiltonian}
\end{equation}
We have subtracted the kinetic operator $t_{\mbox{\tiny c}}$ for
the centre of mass, since it can be eliminated from the problem. The
hamiltonian can be rewritten as
\begin{equation}
H=H_{\mbox{\tiny A}}+H_{\mbox{\tiny B}}+T_{\bm{R}}+V_{\mbox{\tiny AB}},\label{eq:Hamiltonian2}
\end{equation}
where $H_{\mbox{\tiny A}}$ and $H_{\mbox{\tiny B}}$ denote the internal
hamiltonian of each cluster A and B, $T_{\bm{R}}$ denotes the kinetic
operator for the relative motion of the two clusters, and $V_{\mbox{\tiny AB}}$
is the sum of interactions between the two clusters. The wave functions
$\phi_{\mbox{\tiny A}}$ and $\phi_{\mbox{\tiny B}}$ are eigenstates
of $H_{\mbox{\tiny A}}$ and $H_{\mbox{\tiny B}}$ with eigenvalues
$E_{\mbox{\tiny A}}$ and $E_{\mbox{\tiny B}}$, i.e.
\begin{equation}
H_{\mbox{\tiny A}}\phi_{\mbox{\tiny A}}=E_{\mbox{\tiny A}}\phi_{\mbox{\tiny A}}\quad\mbox{and\quad}H_{\mbox{\tiny B}}\phi_{\mbox{\tiny B}}=E_{\mbox{\tiny B}}\phi_{\mbox{\tiny B}}.\label{eq:ClusterEquation}
\end{equation}

There are two ways this can be used to simplify the equation of motion
Eq.~(\ref{eq:EquationOfMotion}). Either one applies the hamiltonian
to the wave functions $\phi_{\mbox{\tiny A}}$ and $\phi_{\mbox{\tiny B}}$
on the right-hand side, or to the wave functions $\phi_{\mbox{\tiny A}}$
and $\phi_{\mbox{\tiny B}}$ on the left-hand side. We refer to these
two equivalent procedures as the RGM1 and RGM2. Although they result
in formally different equations, their solutions are the same.

In the RGM1, one writes
\begin{equation}
\langle\phi_{\mbox{\tiny A}}\phi_{\mbox{\tiny B}}\vert\mathcal{S}[(H-\mathcal{E})\phi_{\mbox{\tiny A}}\phi_{\mbox{\tiny B}}\psi]\rangle=0\label{eq:EquationOfMotion2}
\end{equation}

and using Eqs.~(\ref{eq:Hamiltonian2}) and (\ref{eq:ClusterEquation})
\begin{equation}
\langle\phi_{\mbox{\tiny A}}\phi_{\mbox{\tiny B}}\vert\mathcal{S}[(T_{\bm{R}}+V_{\mbox{\tiny AB}}-E)\phi_{\mbox{\tiny A}}\phi_{\tiny\mbox{B}}\psi]\rangle=0,\label{eq:RGM1-a}
\end{equation}
where $E=\mathcal{E}-E_{\mbox{\tiny A}}-E_{\mbox{\tiny B}}$ is the
scattering energy between the two clusters. The symmetrisation operator
$\mathcal{S}$ can be written as $\mathcal{S}=\hat{1}+\mathcal{S}^{\prime}$,
i.e. the action of $\mathcal{S}$ gives one term leaving the wave
function unchanged, and other terms where particles are exchanged.
Thus, Eq.~(\ref{eq:RGM1-a}) can be written as
\begin{equation}
\boxed{(1-K)\cdot(T_{\bm{R}}-E)\psi+V_{\tiny\mbox{D}}\psi+V_{\tiny\mbox{EX1}}\cdot\psi=0}\label{eq:RGM1}
\end{equation}
where we have introduced a local potential $V_{\mbox{\tiny D}}$ called
the direct potential,
\begin{equation}
V_{\tiny\mbox{D}}(\bm{R})=\langle\phi_{\mbox{\tiny A}}\phi_{\mbox{\tiny B}}\vert V_{\mbox{\tiny AB}}\vert\phi_{\mbox{\tiny A}}\phi_{\mbox{\tiny B}}\rangle,\label{eq:DirectPotential}
\end{equation}
a non-local potential $V_{\mbox{\tiny EX1}}$ called the exchange
potential,
\begin{eqnarray}
V_{\mbox{\tiny EX1}}\cdot\psi & = & \int d^{3}\bm{R}^{\prime}V_{\mbox{\tiny EX1}}(\bm{R},\bm{R}^{\prime})\psi(\bm{R}^{\prime})\label{eq:ExchangePotential1}\\
 & = & \langle\phi_{\mbox{\tiny A}}\phi_{\tiny\mbox{B}}\vert\mathcal{S}^{\prime}[V_{\tiny\mbox{AB}}\phi_{\tiny\mbox{A}}\phi_{\tiny\mbox{B}}\psi]\rangle,\nonumber 
\end{eqnarray}
and a non-local operator $K$ called the exchange kernel,
\begin{eqnarray}
K\cdot\psi & = & \int d^{3}\vec{R}^{\prime}K(\bm{R},\bm{R}^{\prime})\psi(\bm{R}^{\prime})\label{eq:ExchangeKernel}\\
 & = & -\langle\phi_{\tiny\mbox{A}}\phi_{\tiny\mbox{B}}\vert\mathcal{S}^{\prime}[\phi_{\tiny\mbox{A}}\phi_{\tiny\mbox{B}}\psi]\rangle.\nonumber 
\end{eqnarray}

In the RGM2, one applies the hamiltonian Eq.~(\ref{eq:Hamiltonian2})
to the wave functions $\phi_{\tiny\mbox{A}}\phi_{\tiny\mbox{B}}$
on the left-hand side of Eq.~(\ref{eq:EquationOfMotion}), using
Eq.~(\ref{eq:ClusterEquation}). This gives
\begin{equation}
\langle\phi_{\tiny\mbox{A}}\phi_{\tiny\mbox{B}}\vert(T_{\bm{R}}+V_{\tiny\mbox{AB}}-E)\mathcal{S}[\phi_{\tiny\mbox{A}}\phi_{\tiny\mbox{B}}\psi]\rangle=0,\label{eq:RGM2-a}
\end{equation}
which can be written as
\begin{equation}
\boxed{(T_{\bm{R}}-E)(1-K)\cdot\psi+V_{\mbox{\tiny D}}\psi+V_{\mbox{\tiny EX2}}\cdot\psi=0}\label{eq:RGM2}
\end{equation}
where the exchange potential $V_{\mbox{\tiny EX2}}$ is defined by
\begin{eqnarray}
V_{\mbox{\tiny EX2}}\cdot\psi & = & \int d^{3}\vec{R}^{\prime}V_{\mbox{\tiny EX2}}(\vec{R},\vec{R}^{\prime})\psi(\vec{R}^{\prime})\nonumber \\
 & = & \langle\phi_{\tiny\mbox{A}}\phi_{\tiny\mbox{B}}\vert V_{\tiny\mbox{AB}}\mathcal{S}^{\prime}[\phi_{\tiny\mbox{A}}\phi_{\tiny\mbox{B}}\psi]\rangle.\label{eq:ExchangePotential2}
\end{eqnarray}

Hence, the RGM consists in calculating the potentials $V_{\tiny\mbox{D}}$,
$V_{\mbox{\tiny EX}}$ and kernel $K$, and solving the equation for
the relative motion between the two clusters, either Eq.~(\ref{eq:RGM1})
or (\ref{eq:RGM2}). This is of course a great simplification over
solving the full $N$-body equation. Nonetheless, the determination
of $V_{\mbox{\tiny D}}$, $V_{\mbox{\tiny EX}}$ and $K$ involve
$3(n-1)+3(N-n-1)=3(N-2)$-dimensional integrals whose computation
may be costly for large $N$.

\subsection{RGM with contact interactions}

In the following, we apply the RGM to the scattering of universal
clusters. Their universal character is described by the zero-range
theory, which corresponds to the limit of the range of interaction
being much smaller than the $s$-wave scattering length $a$. In this
limit, the interaction potential $V_{ij}$ between two particles appearing
in Eq.~(\ref{eq:Hamiltonian}) and included in the term $V_{\tiny\mbox{AB}}$
in Eq.~(\ref{eq:Hamiltonian2}) can be approximated by a contact
potential,
\begin{equation}
V_{ij}(\bm{r})=g\delta^{3}(\bm{r})\frac{\partial}{\partial r}r\cdot\label{eq:ContactPotential}
\end{equation}
with the coupling constant 
\begin{equation}
g=\frac{4\pi\hbar^{2}a}{2\mu}.\label{eq:CouplingConstant}
\end{equation}
Here, $\mu$ is the reduced mass of the two interacting particles,
and $\partial/\partial r\, r\cdot$ is an operator regularising the
$1/r$ divergence of the wave function when particles come into contact
($r=0$). This potential binds two particles only for $a>0$, and
we restrict our consideration to this case throughout this paper.

The presence of the three-dimensional Dirac delta function in the
potential Eq.~(\ref{eq:ContactPotential}) reduces by three the dimensionality
of the integrals. The dimensionality of the integrals Eqs.~(\ref{eq:DirectPotential}),
(\ref{eq:ExchangePotential1}), and (\ref{eq:ExchangePotential2}),
for $V_{\mbox{\tiny D}}$, $V_{\mbox{\tiny EX1}}$ and $V_{\tiny\mbox{EX2}}$,
is thus reduced to $3(N-3)$.

\subsection{Partial-wave expansion\label{sub:Partial-wave-expansion}}

To proceed further, one can perform a partial-wave expansion in spherical
harmonics $Y_{\ell m}$ in the RGM1 and RGM2 equations. The relative
wave function is expanded as 
\begin{equation}
\psi(\bm{R})=\sum_{\ell m}\frac{1}{R}\psi_{\ell m}(R)Y_{\ell m}(\hat{R}),\label{eq:PartialWaveExpansion}
\end{equation}
where $\hat{R}$ denotes the orientation of $\bm{R}$. Then, the RGM1
equation, Eq.~(\ref{eq:RGM1}), becomes the following set of coupled
equations:
\begin{eqnarray}
\left(T_{R}^{\ell}-E\right)\psi_{\ell m}(R)+\sum_{\ell^{\prime}m^{\prime}}V_{\mbox{\tiny D}}^{\ell m,\ell^{\prime}m^{\prime}}(R)\psi_{\ell^{\prime}m^{\prime}}(R)\label{eq:RGM1PartialWave}\\
-\sum_{\ell^{\prime}m^{\prime}}\int_{0}^{\infty}\!\!\! dR^{\prime}\;\; K^{\ell m,\ell^{\prime}m^{\prime}}(R,R^{\prime})\,(T_{R^{\prime}}^{\ell^{\prime}}-E)\psi_{\ell^{\prime}m^{\prime}}(R^{\prime})\nonumber \\
+\sum_{\ell^{\prime}m^{\prime}}\int_{0}^{\infty}\!\!\! dR^{\prime}\;\; V_{\mbox{\tiny EX1}}^{\ell m,\ell^{\prime}m^{\prime}}(R,R^{\prime})\,\psi_{\ell^{\prime}m^{\prime}}(R^{\prime}) & = & 0\nonumber 
\end{eqnarray}
and the RGM2 equation, Eq.~(\ref{eq:RGM2}), becomes the set of coupled
equations
\begin{eqnarray}
\left(T_{R}^{\ell}-E\right)\psi_{\ell m}(R)+\sum_{\ell^{\prime}m^{\prime}}V_{\mbox{\tiny D}}^{\ell m,\ell^{\prime}m^{\prime}}(R)\psi_{\ell^{\prime}m^{\prime}}(R)\label{eq:RGM2PartialWave}\\
-\sum_{\ell^{\prime}m^{\prime}}\int_{0}^{\infty}\!\!\! dR^{\prime}\;\;(T_{R}^{\ell}-E)K^{\ell m,\ell^{\prime}m^{\prime}}(R,R^{\prime})\,\psi_{\ell^{\prime}m^{\prime}}(R^{\prime})\nonumber \\
+\sum_{\ell^{\prime}m^{\prime}}\int_{0}^{\infty}\!\!\! dR^{\prime}\;\; V_{\mbox{\tiny EX2}}^{\ell m,\ell^{\prime}m^{\prime}}(R,R^{\prime})\,\psi_{\ell^{\prime}m^{\prime}}(R^{\prime}) & = & 0\nonumber 
\end{eqnarray}
with the kinetic energy operator
\begin{equation}
T_{R}^{\ell}=\frac{\hbar^{2}}{2\mu_{N}}\left(-\frac{d^{2}}{dR^{2}}+\frac{\ell(\ell+1)}{R^{2}}\right),\label{eq:KineticOperatorPartialWave}
\end{equation}
where $\mu_{N}$ is the reduced mass of the two clusters, and
\begin{equation}
V_{\mbox{\tiny D}}^{\ell m,\ell^{\prime}m^{\prime}}(R)=\int\!\! d^{2}\hat{R}\;\; Y_{\ell m}^{*}(\hat{R})V_{\mbox{\tiny D}}(\bm{R})Y_{\ell^{\prime}m^{\prime}}(\hat{R}),\label{eq:VDPartialWave}
\end{equation}
\begin{eqnarray}
K^{\ell m,\ell^{\prime}m^{\prime}}(R,R^{\prime}) & = & RR^{\prime}\int\!\! d^{2}\hat{R}d^{2}\hat{R}^{\prime}\;\;\label{eq:KPartialWave}\\
 &  & \quad Y_{\ell m}^{*}(\hat{R})K(\bm{R},\bm{R}^{\prime})Y_{\ell^{\prime}m^{\prime}}(\hat{R}^{\prime}),\nonumber 
\end{eqnarray}
\begin{eqnarray}
V_{\mbox{\tiny EX}}^{\ell m,\ell^{\prime}m^{\prime}}(R,R^{\prime}) & = & RR^{\prime}\int\!\! d^{2}\hat{R}d^{2}\hat{R}^{\prime}\;\;\label{eq:VEXPartialWave}\\
 &  & \quad Y_{\ell m}^{*}(\hat{R})V_{\mbox{\tiny EX}}(\bm{R},\bm{R}^{\prime})Y_{\ell^{\prime}m^{\prime}}(\hat{R}^{\prime}).\nonumber 
\end{eqnarray}

The dimensionality of integration in Eqs.~(\ref{eq:VDPartialWave}),
(\ref{eq:KPartialWave}) and (\ref{eq:VEXPartialWave}) is, respectively,
$3(N-2)-1$, $3(N-2)+1$, and $3(N-2)-2$.

\subsection{Local approximation}

It turns out, as we shall see in the cases treated below, that the
contribution from the non-local kernel $K$ is often small and may
be neglected. Moreover, in some cases, the exchange potentials $V_{\mbox{\tiny EX1}}^{\ell m,\ell^{\prime}m^{\prime}}$
and $V_{\mbox{\tiny EX2}}^{\ell m,\ell^{\prime}m^{\prime}}$ are nearly
local and may be approximated by the local potentials
\begin{equation}
V_{\mbox{\tiny EX local}}^{\ell m,\ell^{\prime}m^{\prime}}(R)=\int_{0}^{\infty}dR^{\prime}V_{\mbox{\tiny EX}}^{\ell m,\ell^{\prime}m^{\prime}}(R,R^{\prime}).\label{eq:localExchangePotential}
\end{equation}

Neglecting $K$ and using the local form Eq.~(\ref{eq:localExchangePotential})
of the exchange potentials constitute the local RGM approximation.
In this approximation, RGM1 and RGM2 equations have the form of conventional
Schr{\"o}dinger equations:
\begin{equation}
\boxed{\left(T_{R}^{\ell}-E\right)\psi_{\ell m}(R)+\sum_{\ell^{\prime}m^{\prime}}V_{1}^{\ell m,\ell^{\prime}m^{\prime}}(R)\psi_{\ell^{\prime}m^{\prime}}(R)=0}\label{eq:localRGM1}
\end{equation}
\begin{equation}
\boxed{\left(T_{R}^{\ell}-E\right)\psi_{\ell m}(R)+\sum_{\ell^{\prime}m^{\prime}}V_{2}^{\ell m,\ell^{\prime}m^{\prime}}(R)\psi_{\ell^{\prime}m^{\prime}}(R)=0}\label{eq:localRGM2}
\end{equation}
where $V_{1}^{\ell m,\ell^{\prime}m^{\prime}}=V_{\mbox{\tiny D}}^{\ell m,\ell^{\prime}m^{\prime}}+V_{\mbox{\tiny EX1 local}}^{\ell m,\ell^{\prime}m^{\prime}}$,
and $V_{2}^{\ell m,\ell^{\prime}m^{\prime}}=V_{\mbox{\tiny D}}^{\ell m,\ell^{\prime}m^{\prime}}+V_{\mbox{\tiny EX2 local}}^{\ell m,\ell^{\prime}m^{\prime}}$.

Unlike the RGM1 and RGM2 equations, Eqs.~(\ref{eq:RGM1PartialWave})
and (\ref{eq:RGM2PartialWave}), the local RGM1 and RGM2 equations,
Eq.~(\ref{eq:localRGM1}) and (\ref{eq:localRGM2}), are not equivalent.
Nevertheless, they often yield similar results as we shall see in
the following sections.

\subsection{Scattering length and scattering volume}

After solving the RGM equations in partial waves, Eq.~(\ref{eq:RGM1PartialWave})
or (\ref{eq:RGM2PartialWave}), or their local-potential approximation,
Eq.~(\ref{eq:localRGM1}) or (\ref{eq:localRGM2}), one obtains the
partial-wave components $\psi_{\ell m}(R)$ of the relative wave function
$\psi$. For zero scattering energy ($E=0$), one can extract the
partial-wave scattering lengths from these components.

From the $s$-wave component $\psi_{00}(R)\propto R+O(1)$ for $R\to\infty$,
one obtains the $s$-wave scattering length,
\begin{equation}
a=\lim_{R\to\infty}R-\frac{\psi_{00}(R)}{\psi_{00}^{\prime}(R)},\label{eq:ScatteringLength}
\end{equation}
and from the $p$-wave component $\psi_{1m}(R)\propto R^{2}+O(1/R)$,
one obtains the $p$-wave scattering volume,
\begin{equation}
v=\lim_{R\to\infty}\frac{R^{3}}{3}\frac{R\psi_{1m}^{\prime}(R)-2\psi_{1m}(R)}{R\psi_{1m}^{\prime}(R)+\psi_{1m}(R)}.\label{eq:ScatteringVolume}
\end{equation}
These formulas follow from the standard definition of the scattering
phase shifts~\cite{Calogero1967}.

\section{Scattering of universal dimers\label{sec:Scattering-of-dimers}}

\subsection{Universal dimers}

We consider universal dimers made of a polarised fermion of mass $M$
and a polarised fermion of mass $m$. These dimers are two-body $s$-wave
weakly-bound states. The normalised wave function $\varphi(r)$ for
the relative motion of the two particles inside the dimer is given
by
\begin{equation}
\varphi(r)=\frac{e^{-r/a}}{\sqrt{2\pi a}r}.\label{eq:DimerWaveFunction}
\end{equation}

\begin{figure}
\includegraphics[scale=0.6]{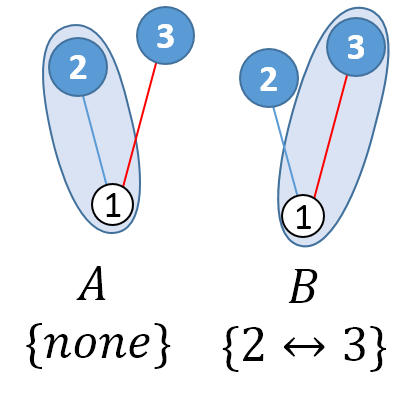}\caption{\label{fig:PermutationsFermionDimer}Schematic representation of the
two permutations of identical fermions between a dimer and a fermion.}

\end{figure}

\subsection{Scattering of a dimer and a particle}

First, we consider the scattering of a universal dimer with a fermionic
particle of mass $M$. To apply the RGM to this case, we set $\phi_{\tiny\mbox{A}}=\varphi$
given by Eq.~(\ref{eq:DimerWaveFunction}), $\phi_{\mbox{\tiny B}}=1$,
and the interaction potential given by Eq.~(\ref{eq:ContactPotential}),
assuming that there is no interaction between identical fermions.
The antisymmetrisation operator in the calculation of the exchange
potentials and kernel is obtained by considering all possible permutations
of identical fermions. In this case, there are two possibilities,
as shown in Fig.~\ref{fig:PermutationsFermionDimer}: no permutation
and the exchange of two fermions of mass $M$. It follows that the
direct and exchange potentials of the RGM equations, Eqs.~(\ref{eq:RGM1}),
and (\ref{eq:RGM2}), are given by the following expressions:
\begin{equation}
V_{\mbox{\tiny D}}(\bm{R})=g\left(\frac{\kappa+1}{\kappa}\right)^{3}\left|\varphi\left(\frac{\kappa+1}{\kappa}R\right)\right|^{2}\label{eq:FermionDimerVD}
\end{equation}
\begin{equation}
V_{\mbox{\tiny EX1}}\cdot\psi(\bm{R})=g\bar{\varphi}^{*}(0)\varphi(R)\psi\Big(-\frac{\kappa}{\kappa+1}\bm{R}\Big)\label{eq:FermionDimerVEX1}
\end{equation}
\begin{equation}
V_{\mbox{\tiny EX2}}\cdot\psi(\bm{R})=g\bar{\varphi}(0)\left(\frac{\kappa+1}{\kappa}\right)^{3}\varphi^{*}\Big(\frac{\kappa+1}{\kappa}R\Big)\psi\Big(-\frac{\kappa+1}{\kappa}\bm{R}\Big),\label{eq:FermionDimerVEX2}
\end{equation}
where $\kappa=M/m$ is the mass ratio and

\begin{equation}
\bar{\varphi}(0)=-\lim_{r\to0}\frac{\partial}{\partial r}r\cdot\varphi(\vec{r})=\frac{1}{\sqrt{2\pi a}a}>0.\label{eq:DimerZeroValue}
\end{equation}

The exchange kernel is given by
\begin{equation}
K\cdot\psi(\vec{R})=\frac{(\kappa+1)^{6}}{(1+2\kappa)^{3}}\int d^{3}\bm{R}^{\prime}\varphi^{*}(\bm{r}_{1})\varphi(\bm{r}_{2})\psi(\bm{R}^{\prime}),\label{eq:FermionDimerK}
\end{equation}
with
\begin{eqnarray*}
\bm{r}_{1} & = & \frac{(\kappa+1)^{2}}{1+2\kappa}\bm{R}^{\prime}+\frac{\kappa(\kappa+1)}{1+2\kappa}\bm{R},\\
\bm{r}_{2} & = & \frac{\kappa(\kappa+1)}{1+2\kappa}\bm{R}^{\prime}+\frac{(\kappa+1)^{2}}{1+2\kappa}\bm{R}.
\end{eqnarray*}

The kinetic operator in Eqs.~(\ref{eq:RGM1}), (\ref{eq:RGM2}) is
given by
\[
T_{\bm{R}}=-\frac{\hbar^{2}}{2(\frac{1}{M+m}+\frac{1}{M})^{-1}}\nabla_{\bm{R}}^{2}.
\]

The RGM1 and RGM2 equations can be solved by performing the partial-wave
expansion of section~\ref{sub:Partial-wave-expansion}. Here, the
potentials Eqs.~(\ref{eq:FermionDimerVD}-\ref{eq:FermionDimerVEX2})
do not couple partial waves:
\begin{equation}
V_{\mbox{\tiny D}}^{\ell m,\ell^{\prime}m^{\prime}}\propto\delta_{\ell,\ell^{\prime}}\delta_{m,m^{\prime}}\label{eq:PartialWaveMatrixElementVD}
\end{equation}
\begin{equation}
V_{\mbox{\tiny EX}}^{\ell m,\ell^{\prime}m^{\prime}}\propto\delta_{\ell,\ell^{\prime}}\delta_{m,m^{\prime}}\label{eq:PartialWaveMatrixElement}
\end{equation}
and for a given partial wave $(\ell,m)$, we obtain from Eq.~(\ref{eq:VEXPartialWave}),
\begin{eqnarray}
\!\!\!\!\!\!\! V_{\mbox{\tiny EX1}}^{\ell m,\ell m}(R,R^{\prime}) & = & (-1)^{\ell}g\frac{\kappa+1}{\kappa}\bar{\varphi}^{*}(0)\label{eq:PartialWaveVEX1}\\
 &  & \times\varphi(R)\delta\Big(R^{\prime}-\!\!\begin{array}{c}
\frac{\kappa}{\kappa+1}\end{array}\!\! R\Big),\nonumber \\
\!\!\!\!\!\!\! V_{\mbox{\tiny EX2}}^{\ell m,\ell m}(R,R^{\prime}) & = & (-1)^{\ell}g\left(\frac{\kappa+1}{\kappa}\right)^{2}\bar{\varphi}(0)\label{eq:PartialWaveVEX2}\\
 &  & \times\varphi^{*}(\!\!\begin{array}{c}
\frac{\kappa+1}{\kappa}\end{array}\!\! R)\delta\Big(R^{\prime}-\!\!\begin{array}{c}
\frac{\kappa+1}{\kappa}\end{array}\!\! R\Big).\nonumber 
\end{eqnarray}

The factor $(-1)^{\ell}$ in these expressions comes from the minus
sign in the argument of $\psi$ in Eqs.~(\ref{eq:FermionDimerVEX1}-\ref{eq:FermionDimerVEX2}).
Due to this factor, the exchange potential is repulsive for even partial
waves, and it is attractive for odd partial waves. Moreover, Eqs.~(\ref{eq:PartialWaveVEX1}-\ref{eq:PartialWaveVEX2})
show that the exchange potentials have an increasingly local character
as the mass ratio $\kappa$ increases. Their local approximation,
given by Eq.~(\ref{eq:localExchangePotential}), leads to
\begin{eqnarray}
\!\!\!\!\!\!\!\!\!\!\!\! V_{\mbox{\tiny EX1 local}}^{\ell m,\ell m}(R) & = & (-1)^{\ell}g\frac{\kappa+1}{\kappa}\bar{\varphi}^{*}(0)\varphi(R),\label{eq:FermionDimerVEX1Local}\\
\!\!\!\!\!\!\!\!\!\!\!\! V_{\mbox{\tiny EX2 local}}^{\ell m,\ell m}(R) & = & (-1)^{\ell}g\left(\frac{\kappa+1}{\kappa}\right)^{2}\bar{\varphi}(0)\varphi^{*}\Big(\!\!\begin{array}{c}
\frac{\kappa+1}{\kappa}\end{array}\!\! R\Big).\label{eq:FermionDimerVEX2Local}
\end{eqnarray}

We solve the resulting RGM and local RGM equations numerically by
discretising the coordinate~$R$.

\subsubsection{s-wave scattering}

\begin{figure}
\includegraphics[scale=0.6]{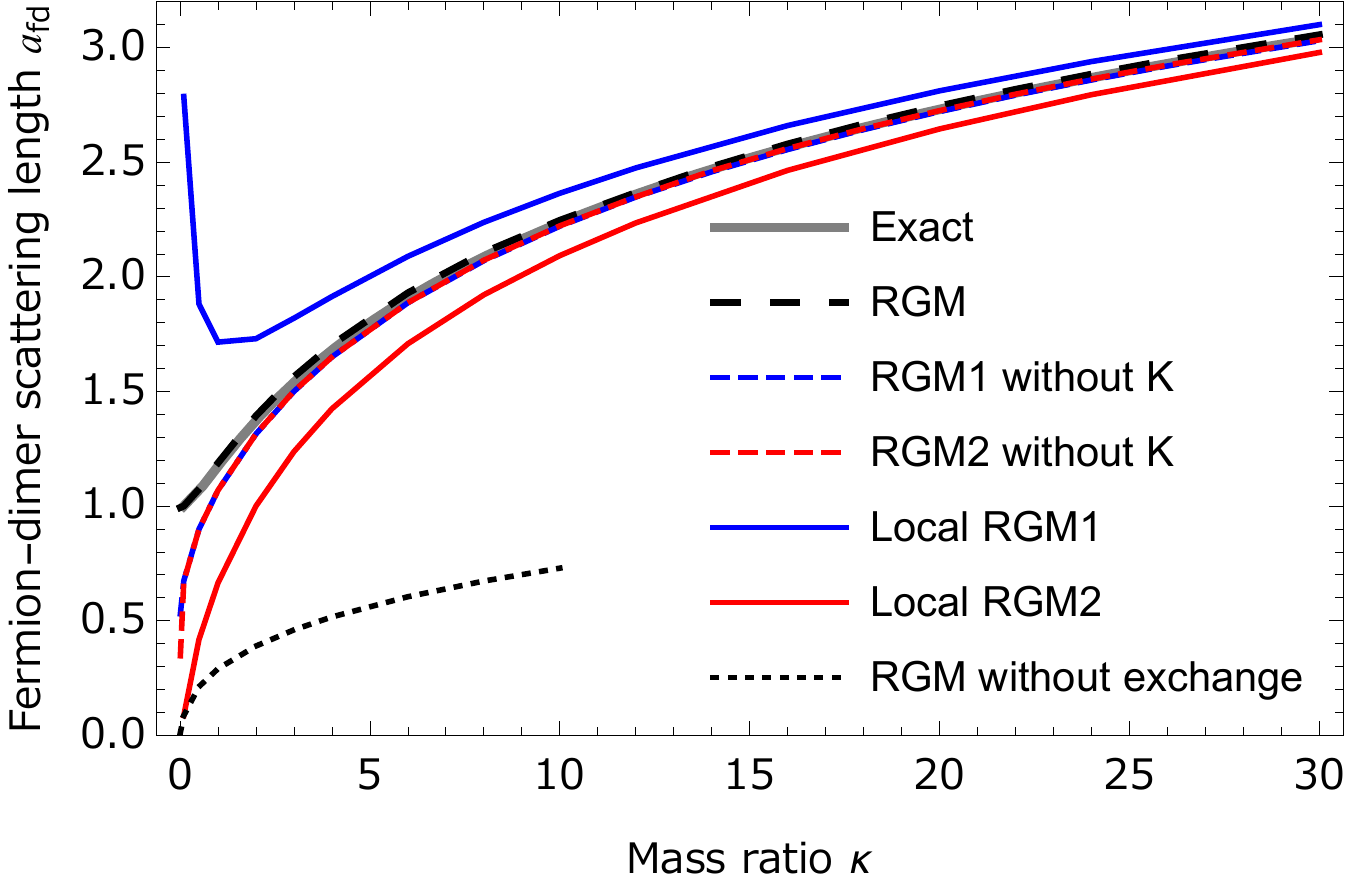}

\caption{\label{fig:Fermion-dimer-scattering-length}Fermion-dimer $s$-wave
scattering length $a_{fd}$ in units of $a$ as a function of the
mass ratio.}
\end{figure}

We first consider fermion-dimer scattering in the $s$~wave, for
which the effective potential is purely repulsive. The fermion-dimer
$s$-wave scattering length $a_{fd}$ is therefore always positive.
It is shown in Fig.~\ref{fig:Fermion-dimer-scattering-length}, as
a function of the mass ratio $\kappa$. For the equal mass case ($M=m$),
we obtain
\[
a_{fd}=1.19a,
\]
which is consistent with the exact result $\approx1.17907a$~\cite{Skorniakov1957,Iskin2010}.
All RGM results are within 2\% of the exact results, indicating that
there is little excitation during the collision of a dimer and fermion,
the dimer remaining bound during the collision. Nonetheless, the exchange
of particles is crucial. The dotted curve in Fig.~\ref{fig:Fermion-dimer-scattering-length}
shows that including only the direct potential (neglecting the exchange
kernel and potential) yields a much smaller scattering length. On
the other hand, the exchange kernel $K$ brings a significant difference
only for mass ratios smaller than one, and may be neglected otherwise,
as shown by the dashed red and blue curves in Fig.~\ref{fig:Fermion-dimer-scattering-length}.
As to the local approximation, it leads to results which are close
to those of the RGM for sufficiently large mass ratios, as seen from
the red and blue curves in Fig.~\ref{fig:Fermion-dimer-scattering-length}.

\subsubsection{p-wave scattering}

\begin{figure}
\includegraphics[scale=0.58]{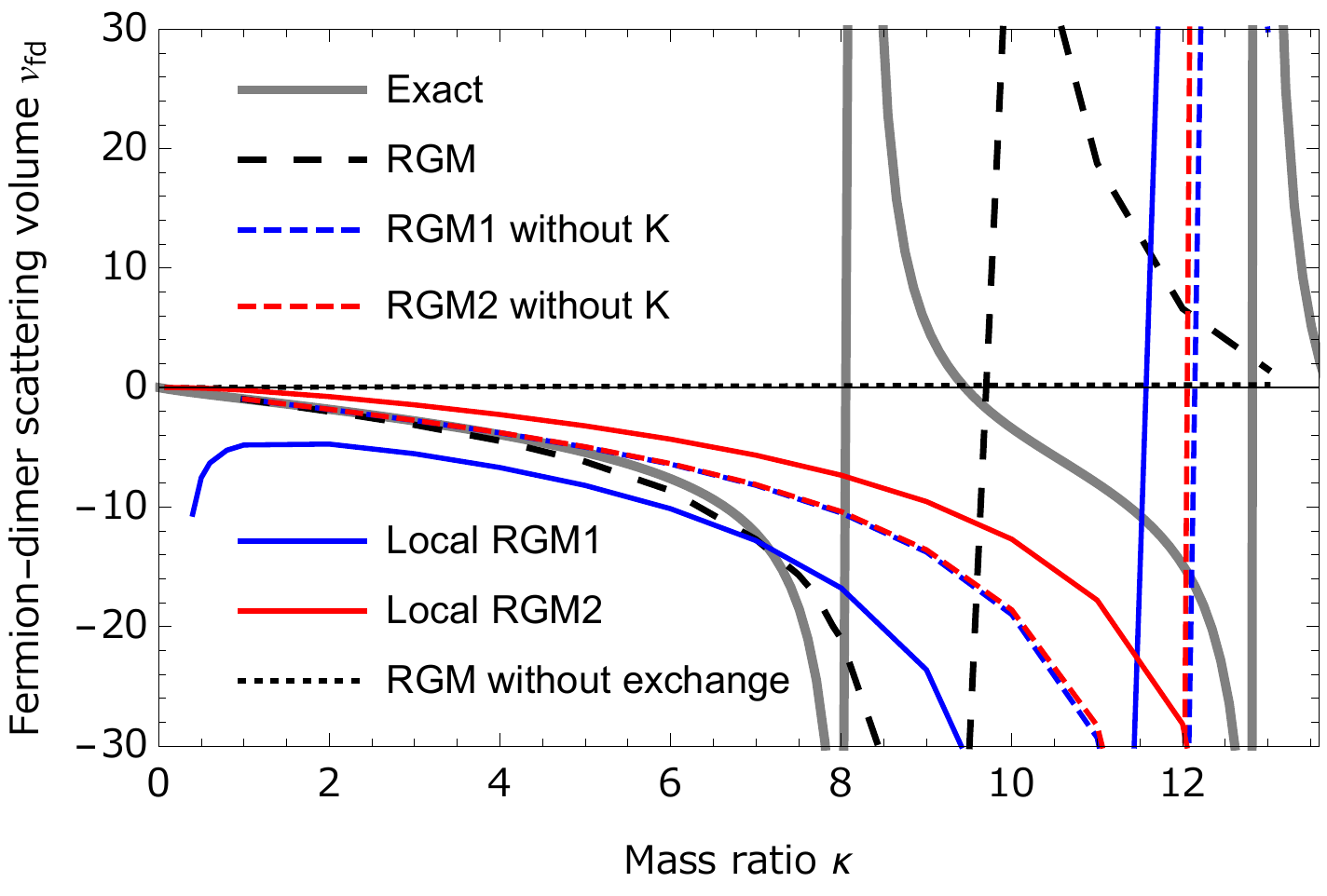}

\caption{\label{fig:Fermion-dimer-scattering-length-P-wave}Fermion-dimer $p$-wave
scattering volume $v_{fd}$ in units of $a^{3}$ as a function of
the mass ratio.}
\end{figure}

In the $p$-wave channel, the fermion and dimer attract each other.
This is due to the Efimov attraction~\cite{Efimov1970a,Efimov1973}
that results from the effective interaction between the two heavy
fermions mediated by the light fermion. Although the Efimov attraction
wins over the centrifugal barrier only for mass ratios $M/m>\kappa_{c}\approx13.6069657$
\cite{Efimov1973,Petrov2003,Kartavtsev2007}, resulting in an infinite
discrete-scale-invariant tower of three-body bound states, it also
makes the system attractive for lower mass ratios, resulting in an
overall negative $p$-wave scattering length. As the mass ratio increases,
the Efimov attraction strengthens, and two universal three-body bound
states appear at mass ratios $\kappa_{1}=8.17260$ and $\kappa_{2}=12.91743$~\cite{Kartavtsev2007}.
At these mass ratios, fermion-dimer $p$-wave scattering is resonant
and the $p$-wave scattering volume $v_{fd}$ diverges, as shown in
Fig.~\ref{fig:Fermion-dimer-scattering-length-P-wave}.

In the RGM, the effective potential between the fermion and the dimer
scattering in the $p$ wave is also attractive, due to the factor
$(-1)^{\ell}$ of Eqs.~(\ref{eq:PartialWaveVEX1}-\ref{eq:PartialWaveVEX2}).
The scattering volume calculated in the RGM is thus negative and very
close to the exact one up to the mass ratio $M/m\approx6$. For the
equal mass case ($M=m$), the RGM gives
\[
v_{fd}=-0.98a,
\]
which is consistent with the exact result $\approx-0.96a$~\cite{Endo2011}.
Beyond the mass ratio $\sim6$, the RGM results deviate strongly from
the exact results. This is explained by the fact that the resonance
and the three-body bound state at $M/m=\kappa_{1}$ imply three-body
correlations that are not fully captured by the RGM. Nevertheless,
the RGM exhibits a similar resonance, but at a shifted mass ratio
$\kappa_{1}^{\mbox{\tiny(RGM)}}\approx9.5$. This shows that the Efimov
attraction, physically due to the exchange of light fermion between
the two heavy fermions, is partially captured by the mere antisymmetrisation
of the wave function in the RGM, as suggested by Fig.~\ref{fig:PermutationsFermionDimer}.

The local RGM equations reproduce approximately the RGM results for
$M/m<6$, as shown by the blue and red curves in Fig.~\ref{fig:Fermion-dimer-scattering-length-P-wave}.
For larger mass ratios, the difference between the local RGM and full
RGM results is substantial and it is mainly due to the absence of
the exchange kernel $K$ in the local RGM equations, as shown by the
dashed red and blue curves in Fig.~\ref{fig:Fermion-dimer-scattering-length-P-wave}.

\subsection{Scattering of two dimers}

\begin{figure}
\includegraphics[scale=0.6]{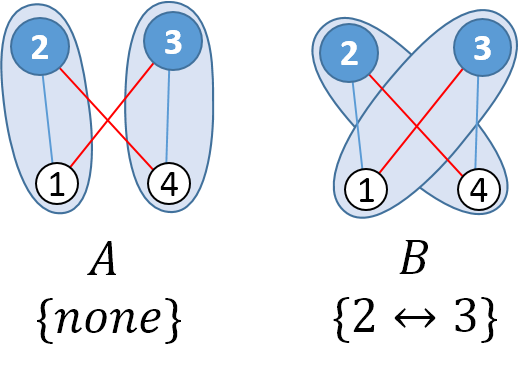}\caption{\label{fig:PermutationsDimerDimer}Schematic representation of the
two permutations of identical fermions between two dimers.}
\end{figure}

Now, we consider the scattering of two universal dimers. We thus apply
the RGM equations for the two cluster wave functions $\phi_{\mbox{\tiny A}}=\phi_{\mbox{\tiny B}}=\varphi$
given by Eq.~(\ref{eq:DimerWaveFunction}) and the interaction potential
given by Eq.~(\ref{eq:ContactPotential}), assuming again that there
is no interaction between identical fermions. The antisymmetrisation
operator in the calculation of the exchange potentials and kernel
is obtained by considering all possible permutations of identical
fermions. In this case, there are two possibilities, as shown in Fig.~\ref{fig:PermutationsDimerDimer}:
no permutation and the exchange of two fermions of mass $M$ (which
is equivalent to exchanging the two fermions of mass $m$). After
some straightforward calculations, the direct and exchange potentials,
as well as the exchange kernel of the RGM equations, Eq.~(\ref{eq:RGM1})
and (\ref{eq:RGM2}), are given by the following expressions:
\[
V_{\mbox{\tiny D}}(\bm{R})=2g\left(\kappa+1\right)^{3}\int d^{3}\bm{r}\vert\varphi(\bm{r})\varphi((\kappa+1)\bm{R}+\kappa\bm{r})\vert^{2},
\]
\[
K(\!\bm{R},\bm{R}^{\prime}\!)\!=\!-\!\lambda\!\int\!\! d^{3}\!\bm{r}\varphi^{*}\!(\bm{r}+\bm{R}_{3})\varphi^{*}\!(\bm{r})\varphi(\bm{r}+\bm{R}_{1})\varphi(\bm{r}+\bm{R}_{2}),
\]
\[
V_{\mbox{\tiny EX1}}(\bm{R},\bm{R}^{\prime})=2g\lambda\varphi^{*}(\bm{R}_{3})\bar{\varphi}^{*}(0)\varphi(\bm{R}_{1})\varphi(\bm{R}_{2}),
\]
\[
V_{\mbox{\tiny EX2}}(\bm{R},\bm{R}^{\prime})=2g\lambda\varphi^{*}(\bm{R}_{2})\varphi^{*}(\bm{R}_{1})\bar{\varphi}(0)\varphi(\bm{R}_{4}).
\]

In these expressions, we have set $\kappa=M/m$, $\lambda=(\kappa+1)^{6}/(2\kappa)^{3}$,
and
\begin{eqnarray*}
\bm{R}_{1} & = & \frac{\kappa+1}{2\kappa}\left(\bm{R}^{\prime}+\bm{R}\right),\\
\bm{R}_{2} & = & \frac{\kappa+1}{2}\left(\bm{R}^{\prime}-\bm{R}\right),\\
\bm{R}_{3} & = & \frac{(\kappa+1)}{2\kappa}\left((\kappa+1)\bm{R}^{\prime}-(\kappa-1)\bm{R}\right),\\
\bm{R}_{4} & = & \frac{\kappa+1}{2\kappa}\left((\kappa-1)\bm{R}^{\prime}-(\kappa+1)\bm{R}\right).
\end{eqnarray*}

The kinetic operator in Eqs.~(\ref{eq:RGM1}), (\ref{eq:RGM2}),
(\ref{eq:localRGM1}) and (\ref{eq:localRGM2}) is given by
\begin{equation}
T_{\bm{R}}=-\frac{\hbar^{2}}{M+m}\vec{\nabla}_{\bm{R}}^{2}.\label{eq:DimerDimerKineticOperator}
\end{equation}
\begin{figure}
\includegraphics[scale=0.6]{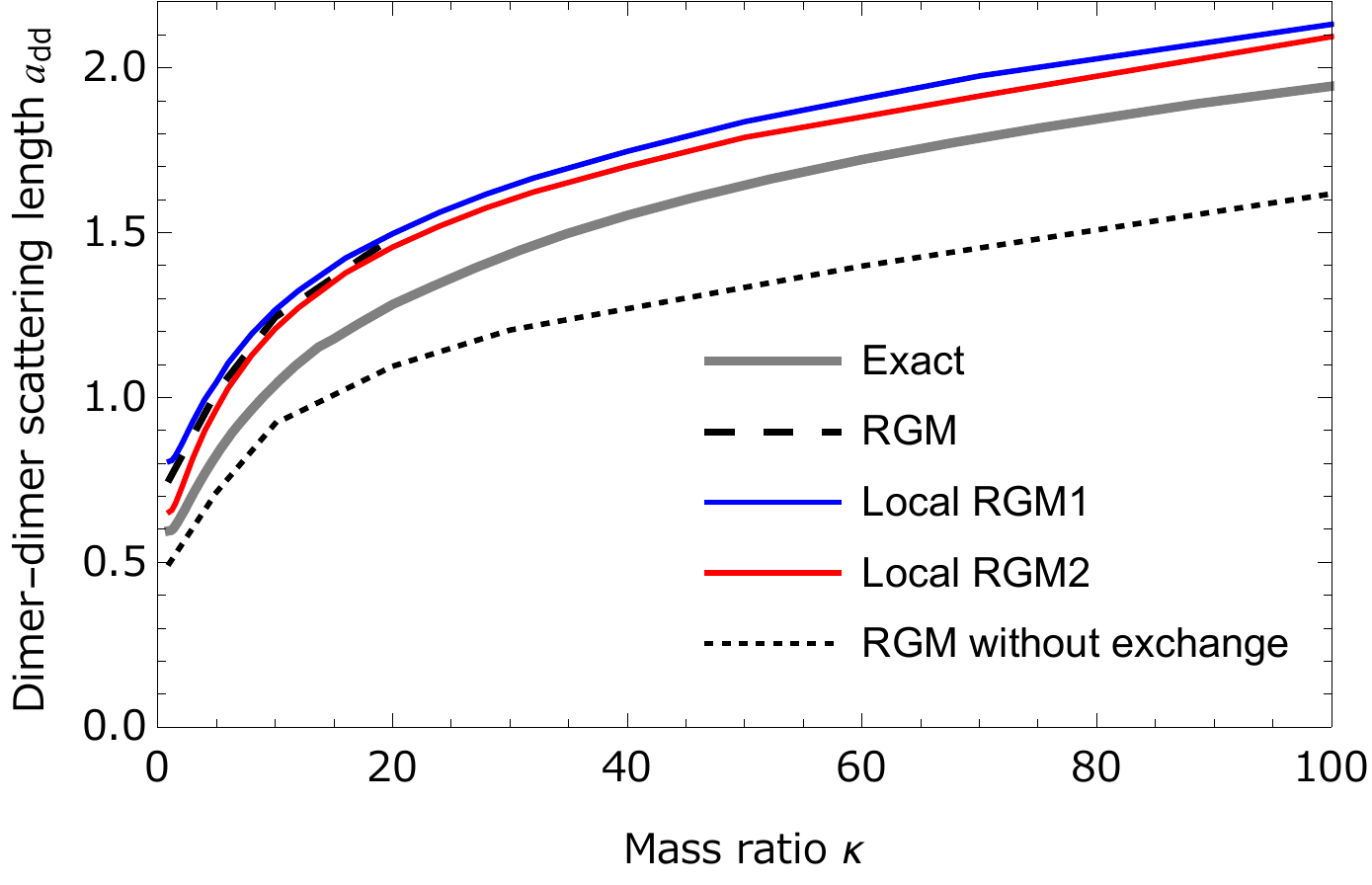}\caption{\label{fig:Dimer-dimer-scattering-length}Dimer-dimer scattering length
$a_{dd}$ in units of $a$ as a function of the mass ratio.}
\end{figure}

We solve the RGM1 and RGM2 equations, as well as their local approximation,
by performing the partial-wave expansion of section~\ref{sub:Partial-wave-expansion}
and discretising the coordinate $R$. The potentials are repulsive
for all partial waves. The resulting dimer-dimer $s$-wave scattering
length is shown in Fig.~\ref{fig:Dimer-dimer-scattering-length},
as a function of the mass ratio $\kappa$. For the equal mass case
($M=m$), we obtain
\[
a_{dd}=0.752a,
\]
which is close, although significantly different, from the exact result
$\approx0.6a$~\cite{Petrov2004a,Astrakharchik2004,Petrov2005,Petrov2005a,Brodsky2006,Levinsen2006,Stecher2007,Marcelis2008}.
This means that compared to dimer-particle scattering, there is a
bit more excitation during the collision of two dimers, although it
remains small. On the other hand, exchange is less important than
in the case of fermion-dimer scattering, as the major contribution
to the scattering length comes from the direct potential, as seen
from the dotted curve in Fig.~\ref{fig:Dimer-dimer-scattering-length}.
The exchange potentials have an increasingly local character as the
mass ratio increases. We have calculated the scattering length with
the RGM up to mass ratio 20. Beyond this mass ratio, the local character
of the potential makes it difficult to solve the problem as a non-local
one, since a high degree of discretisation is needed. The local RGM
equations, on the other hand, are easier to solve. They give results
which are very close to the RGM, as can be seen from the blue and
red curves of Fig.~\ref{fig:Dimer-dimer-scattering-length}, and
can easily be extended to larger mass ratios.

\section{Scattering of universal trimers\label{sec:Scattering-of-trimers}}

\subsection{Universal trimers}

We now consider universal trimers made of two polarised fermions of
mass $M$ and a polarised fermion of mass $m$. Such trimers exist
for a mass ratio $M/m>\kappa_{1}\approx8.17260$. They rotate with
one quantum unit of angular momentum, and can therefore be in three
possible internal quantum states of rotation, labelled by the quantum
number $m\in\{-1,0,1\}$. For a mass ratio $M/m>\kappa_{c}\approx13.6069657$~\cite{Efimov1973,Petrov2003,Kartavtsev2007},
the trimers are Efimov states~\cite{Efimov1970a,Braaten2006}, characterised
by the scattering length $a$ between the two different kinds of fermions,
and a three-body parameter. For a mass ratio $M/m<\kappa_{c}$, the
trimers are Kartavtsev-Malykh states~\cite{Kartavtsev2007}, characterised
only by the scattering length $a$. We restrict our consideration
to these states, and therefore to the range $\kappa_{1}<M/m<\kappa_{c}$
where a ground-state trimer exists.

\begin{figure}
\includegraphics{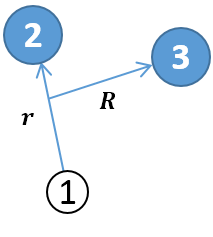}

\caption{\label{fig:Jacobi-coordinates}Jacobi coordinates $\bm{r}$ and $\bm{R}$
decribing a trimer made of two heavy fermions and a light fermion.
The vector $\bm{r}=\bm{r}_{2}-\bm{r}_{1}$ is the relative position
between particle 2 and 1, and the vector $\bm{R}=\bm{r}_{3}-\frac{M\bm{r}_{2}+m\bm{r}_{1}}{M+m}$
is the relative position between particle 3 and the centre of mass
of particles 1 and 2.}

\end{figure}

The trimer wave function is expressed as a function of Jacobi vectors
$\bm{r}$ and $\bm{R}$ shown in Fig.~\ref{fig:Jacobi-coordinates}.
To a good accuracy, the trimer wave function is well approximated
by the adiabatic hyperspherical form~\cite{Kartavtsev2007}

\begin{equation}
\phi_{m}(\bm{r},\bm{R})=\frac{f(\mathcal{R})}{\mathcal{R}^{5/2}}\left[\psi_{\mbox{\tiny Fad}}(\mathcal{R},\alpha,\hat{y})-\psi_{\mbox{\tiny Fad}}(\mathcal{R},\tilde{\alpha},\hat{\tilde{y}})\right],\label{eq:KMtrimer}
\end{equation}
where the component
\begin{equation}
\psi_{\mbox{\tiny Fad}}(\mathcal{R},\alpha,\hat{y})=\frac{C(\mathcal{R})}{\sin2\alpha}\psi_{\mbox{\tiny ang}}(\mathcal{R},\alpha)Y_{1}^{m}(\hat{y})\label{eq:FaddeevComponent}
\end{equation}
incorporates the angular momentum of the trimer through the spherical
harmonic $Y_{1}^{m}$. The hyperangular component $\psi_{\tiny\mbox{ang}}$
is given by
\begin{eqnarray}
\psi_{\tiny\mbox{ang}}(\mathcal{R},\alpha) & = & \cosh\left[s(\mathcal{R})\left(\frac{\pi}{2}-\alpha\right)\right]\label{eq:Psiang}\\
 &  & -\frac{\tan\alpha}{s(\mathcal{R})}\sinh\left[s(\mathcal{R})\left(\frac{\pi}{2}-\alpha\right)\right].\nonumber 
\end{eqnarray}
Here, we use the following hyperspherical coordinates
\begin{eqnarray*}
\mathcal{R} & = & \sqrt{x^{2}+y^{2}}=\sqrt{\tilde{x}^{2}+\tilde{y}^{2}}\\
\bm{x} & = & \beta^{1/2}\bm{r}\qquad\qquad\mbox{with }\beta=\frac{\sqrt{2\kappa+1}}{\kappa+1}\\
\bm{y} & = & \beta^{-1/2}\bm{R}
\end{eqnarray*}
 
\begin{eqnarray*}
\bm{\tilde{x}} & = & \sin\omega\bm{x}+\cos\omega\bm{y}\qquad\mbox{with }\cot\omega=\frac{\sqrt{2\kappa+1}}{\kappa}\\
\bm{\tilde{y}} & = & \cos\omega\bm{x}-\sin\omega\bm{y}
\end{eqnarray*}
\begin{eqnarray*}
\alpha & = & \arctan\left(\frac{x}{y}\right)\\
\tilde{\alpha} & = & \arctan\left(\frac{\tilde{x}}{\tilde{y}}\right)
\end{eqnarray*}

The function $s(\mathcal{R})$ is determined by
\begin{eqnarray*}
\beta^{-1/2}\frac{\mathcal{R}}{a} & = & \frac{1+s^{2}}{s}\tanh\Big(s\frac{\pi}{2}\Big)\\
 &  & -\frac{2}{\sin2\omega}\frac{\cosh s\omega}{\cosh(s\frac{\pi}{2})}+\frac{\sinh s\omega}{s\,\sin^{2}\omega\,\cosh(s\frac{\pi}{2})}.
\end{eqnarray*}

The function $f(\mathcal{R})$ is the solution asssociated with the
lowest eigenvalue $\varepsilon_{\tiny\mbox{trimer}}$ of the hyper-radial
equation
\[
\left[-\frac{d^{2}}{dR^{2}}-\frac{s^{2}(\mathcal{R})+\frac{1}{4}}{\mathcal{R}^{2}}-\varepsilon_{\mbox{\tiny trimer}}\right]f(\mathcal{R})=0,
\]
and normalised as 
\[
\int_{0}^{\infty}d\mathcal{R}\vert f(\mathcal{R})\vert^{2}=1.
\]

The function $C(\mathcal{R})$ is determined by the normalisation
condition
\begin{multline*}
\frac{1}{4}\int(\sin2\alpha)^{2}d\alpha d\Omega_{x}d\Omega_{y}\times\\
\left|\psi_{\mbox{Fad}}(\mathcal{R},\alpha,\hat{y})-\psi_{\mbox{Fad}}(\mathcal{R},\tilde{\alpha},\hat{\tilde{y}})\right|^{2}=1,
\end{multline*}
which guarantees that
\begin{eqnarray*}
 &  & \int d^{3}\bm{r}d^{3}\bm{R}\;\vert\phi(\bm{r},\bm{R})\vert^{2}\\
 & = & \frac{1}{4}\int\mathcal{R}^{5}d\mathcal{R}(\sin2\alpha)^{2}d\alpha d\Omega_{x}d\Omega_{y}\;\vert\phi(\bm{r},\bm{R})\vert^{2}\\
 & = & 1.
\end{eqnarray*}

\subsection{Scattering of two trimers}

Trimers in the same rotational state $i$ are identical fermions and
therefore scatter only in the $p$ wave channel at low-energy. At
sufficiently low energy, this $p$-wave scattering is negligible with
respect to the $s$-wave scattering between trimers in different rotational
states. For this reason, we focus on the latter in this paper. There
are three possible pairs of different rotational states, $\{-1,0\}$,
$\{0,1\}$, and $\{1,-1\}$, and they all lead to the same scattering
length, because of the $\mbox{SU}(3)$ symmetry of this system. However,
this symmetry is artificially broken by the single-channel RGM, if
rotational states are given by the usual spherical harmonics. The
different values of scattering lengths for the different pairs of
states would thus give an indication of the error of the single-channel
RGM approximation. However, a more serious issue is that spherical
harmonics are complex-valued and the RGM does not ensure the scattering
length to be real. To circumvent this problem, we consider an alternative
basis for rotational states, which is the $xyz$ basis formed by rotational
states with angular momentum projection zero on the three axes of
space. Namely,

\[
Y_{1}^{x}=\frac{Y_{1}^{1}-Y_{1}^{-1}}{\sqrt{2}};\;\; Y_{1}^{y}=\frac{Y_{1}^{1}+Y_{1}^{-1}}{i\sqrt{2}};\;\; Y_{1}^{z}=Y_{1}^{0}.
\]
In an exact calculation, it makes no difference whether one uses the
usual spherical harmonics or the $xyz$ basis, but in the case of
the RGM, the $xyz$ basis ensures the results to be real, since the
wave functions in this basis are all real, and restores the $\mbox{SU}(3)$
symmetry as well. This is evident if one observes that the three pairs
$\{xy\}$, $\{yz\}$, and $\{zx\}$ can be transformed into each other
by a rotation in space.

\begin{figure*}
\includegraphics[scale=0.5]{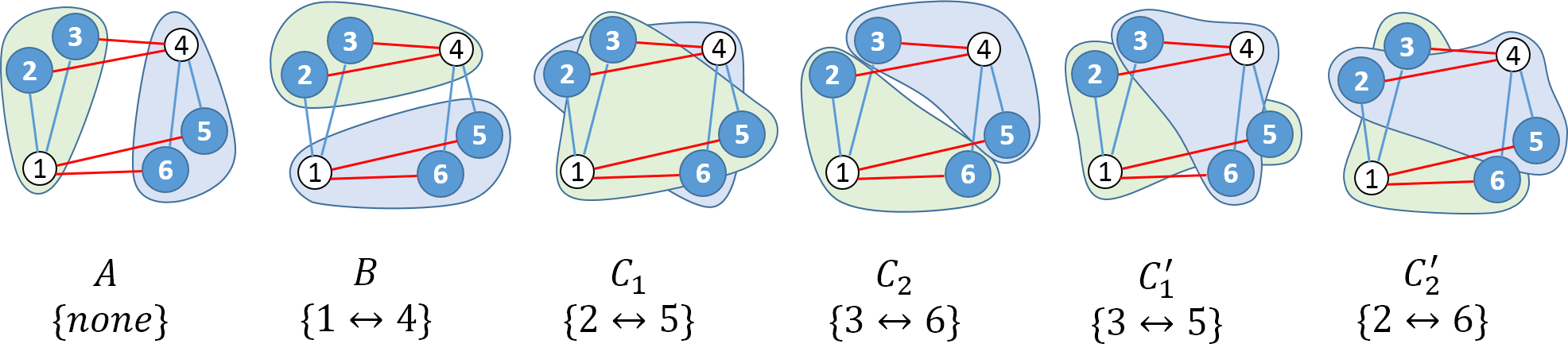}

\caption{\label{fig:PermutationsTrimer}Schematic representation of the first
six permutations of identical fermions between two trimers. The next
six permutations are obtained by performing the permutation $\{1\leftrightarrow4,\,2\leftrightarrow5,\,3\leftrightarrow6\}$
(which produces a minus sign for the corresponding terms in the wave
function).}
\end{figure*}

To apply the RGM to this scattering problem, we set $\phi_{A}=\phi_{x}$
and $\phi_{B}=\phi_{y}$ (i.e. the two clusters are two trimers in
rotational state $x$ and $y$). There are twelve possible permutations
of identical fermions between the two trimers, as shown in Fig.~\ref{fig:PermutationsTrimer}.
From this we obtain the expressions for the direct potential, the
exchange potentials and the exchange kernel, which are given respectively
by the following nine, six, and nine-dimensional integrals:\begin{widetext}
\begin{equation}
V_{D}(\bm{s})=2g\left(\frac{\kappa+1}{\kappa}\right)^{3}\int d^{3}\bm{R}d^{3}\bm{r}^{\prime}d^{3}\bm{R}^{\prime}\left(\left|\phi_{x}(\bm{r}_{-},\bm{R})\phi_{y}(\bm{r}^{\prime},\bm{R}^{\prime})\right|^{2}+\left|\phi_{x}(\bm{r}^{\prime},\bm{R}^{\prime})\phi_{y}(\bm{r}_{+},\bm{R})\right|^{2}\right)\label{eq:TrimerTrimerVD}
\end{equation}
with 
\begin{equation}
\bm{r}_{\pm}=-\frac{\kappa+1}{2\kappa+1}\bm{R}\pm\frac{\kappa+1}{\kappa}\bm{s}+\frac{\kappa+1}{2\kappa+1}\bm{R}^{\prime}-\frac{1}{\kappa}\bm{r}^{\prime}\label{eq:rplusminus}
\end{equation}

\begin{eqnarray}
V_{\mbox{\tiny EX1}}(\bm{s},\bm{s}^{\prime})\!\! & = & g\lambda\int\! d^{3}\bm{r}d^{3}\bm{R}\;\Big(\;\left(\bar{\phi}_{x}(\bm{R}_{1})\phi_{y}(\vec{\mathcal{R}})\mp\bar{\phi}_{y}(\bm{R}_{1})\phi_{x}(\bm{\mathcal{R}})\right)^{*}\left(\phi_{x}(\bm{\mathcal{R}}_{2})\phi_{y}(\bm{\mathcal{R}}_{3})\mp\phi_{y}(\bm{\mathcal{R}}_{2})\phi_{x}(\bm{\mathcal{R}}_{3})\right)\label{eq:TrimerTrimerVEX1}\\
 &  & +\frac{2}{\kappa^{3}}\left(\bar{\phi}_{x}(\bm{R}_{1}^{\prime})\phi_{y}(\bm{\mathcal{R}})\mp\bar{\phi}_{y}(\bm{R}_{1}^{\prime})\phi_{x}(\bm{\mathcal{R}})\right)^{*}\left(\phi_{x}(\bm{\mathcal{R}}_{2}^{\prime})\phi_{y}(\bm{\mathcal{R}}_{3}^{\prime})\mp\phi_{y}(\bm{\mathcal{R}}_{2}^{\prime})\phi_{x}(\bm{\mathcal{R}}_{3}^{\prime})\right)\nonumber \\
 &  & -\frac{2}{\kappa^{3}}\left(\phi_{x}(\bm{\mathcal{R}}_{1}^{\prime\prime})\phi_{y}(\bm{\mathcal{R}})\mp\phi_{y}(\bm{\mathcal{R}}_{1}^{\prime\prime})\phi_{x}(\bm{\mathcal{R}})\right)^{*}\left(\phi_{x}(\bm{\mathcal{R}}_{2}^{\prime\prime})\phi_{y}(\bm{\mathcal{R}}_{3}^{\prime\prime})\mp\phi_{y}(\bm{\mathcal{R}}_{2}^{\prime\prime})\phi_{x}(\bm{\mathcal{R}}_{3}^{\prime\prime})\right)\;\;\Big)\nonumber 
\end{eqnarray}
\begin{eqnarray}
V_{\mbox{\tiny EX2}}(\bm{s},\bm{s}^{\prime})\!\! & = & g\lambda\int\! d^{3}\bm{r}d^{3}\bm{R}\;\Big(\;\left(\phi_{x}(\bm{\mathcal{R}}_{3})\bar{\phi}_{y}(\bm{R}_{4})\mp\phi_{y}(\bm{\mathcal{R}}_{3})\bar{\phi}_{x}(\bm{R}_{4})\right)\left(\phi_{x}(\bm{\mathcal{R}})\phi_{y}(\bm{\mathcal{R}}_{5})\mp\phi_{y}(\bm{\mathcal{R}})\phi_{x}(\bm{\mathcal{R}}_{5})\right)^{*}\label{eq:TrimerTrimerVEX2}\\
 &  & +\frac{2}{\kappa^{3}}\left(\phi_{x}(\bm{\mathcal{R}}_{3}^{\prime})\bar{\phi}_{y}(\bm{R}_{4}^{\prime})\mp\phi_{y}(\bm{\mathcal{R}}_{3}^{\prime})\bar{\phi}_{x}(\bm{R}_{4}^{\prime})\right)\left(\phi_{x}(\bm{\mathcal{R}})\phi_{y}(\bm{\mathcal{R}}_{5}^{\prime})\mp\phi_{y}(\bm{\mathcal{R}})\phi_{x}(\bm{\mathcal{R}}_{5}^{\prime})\right)^{*}\nonumber \\
 &  & -\frac{2}{\kappa^{3}}\left(\phi_{x}(\bm{\mathcal{R}}_{1}^{\prime\prime})\phi_{y}(\bm{\mathcal{R}})\mp\phi_{y}(\bm{\mathcal{R}}_{1}^{\prime\prime})\phi_{x}(\bm{\mathcal{R}})\right)^{*}\left(\phi_{x}(\bm{\mathcal{R}}_{2}^{\prime\prime})\phi_{y}(\bm{\mathcal{R}}_{3}^{\prime\prime})\mp\phi_{y}(\bm{\mathcal{R}}_{2}^{\prime\prime})\phi_{x}(\bm{\mathcal{R}}_{3}^{\prime\prime})\right)\;\;\Big).\nonumber 
\end{eqnarray}
\begin{eqnarray}
K_{\mbox{ }}(\bm{s},\bm{s}^{\prime})\!\! & = & \frac{\lambda}{4}\int d^{3}\bm{r}d^{3}\bm{R}d^{3}\bm{r}^{\prime}\Big(\left(\phi_{x}(\bm{\mathcal{R}}_{3})\phi_{y}(\bm{\mathcal{R}}_{4})\mp\phi_{y}(\bm{\mathcal{R}}_{3})\phi_{x}(\bm{\mathcal{R}}_{4})\right)\left(\phi_{x}(\bm{\mathcal{R}})\phi_{y}(\bm{r}^{\prime},\bm{R}_{5})\mp\phi_{y}(\bm{\mathcal{R}})\phi_{x}(\bm{r}^{\prime},\bm{R}_{5})\right)^{*}\nonumber \\
 &  & +\frac{4}{\kappa^{3}}\left(\phi_{x}(\bm{\mathcal{R}}_{3}^{\prime})\phi_{y}(\bm{\mathcal{R}}_{4}^{\prime})\mp\phi_{y}(\bm{\mathcal{R}}_{3}^{\prime})\phi_{x}(\bm{\mathcal{R}}_{4}^{\prime})\right)\left(\phi_{x}(\bm{\mathcal{R}})\phi_{y}(\bm{r}^{\prime},\bm{R}_{5}^{\prime})\mp\phi_{y}(\bm{\mathcal{R}})\phi_{x}(\bm{r}^{\prime},\bm{R}_{5}^{\prime})\right)^{*}\;\;\Big).\label{eq:TrimerTrimerK}
\end{eqnarray}

{\scriptsize{}}\end{widetext}

In these expressions, we have set
\[
\lambda=2\frac{(2\kappa+1)^{6}}{(2\kappa)^{3}},
\]
\[
\bar{\phi}_{m}(\bm{R})=-\lim_{r\to0}\frac{\partial}{\partial r}r\phi_{m}(\bm{r},\bm{R}),
\]
and $\bm{\mathcal{R}}_{i}$ stands for $(\bm{r}_{i},\bm{R}_{i})$.
These variables are given explicitly in terms of $\bm{r}$ and $\bm{R}$
in the Appendix. The sign $\mp$ in Eqs.~(\ref{eq:TrimerTrimerVEX1}-\ref{eq:TrimerTrimerVEX2})
is $-$ for even scattering waves, and $+$ for odd scattering waves.
Since we are interested in $s$-wave scattering, only even waves are
involved due to the conservation of parity, and thus $\mp=-$ in this
case. Note that the asterisk in Eqs.~(\ref{eq:TrimerTrimerVEX1}-\ref{eq:TrimerTrimerK})
still denotes the complex conjugate, although in our calculations
all wave functions are real. To compute these high-dimensional integrals,
we resort to Monte Carlo integration using importance sampling.

\begin{figure}
\includegraphics[bb=0bp 0bp 575bp 401bp,clip,scale=0.41]{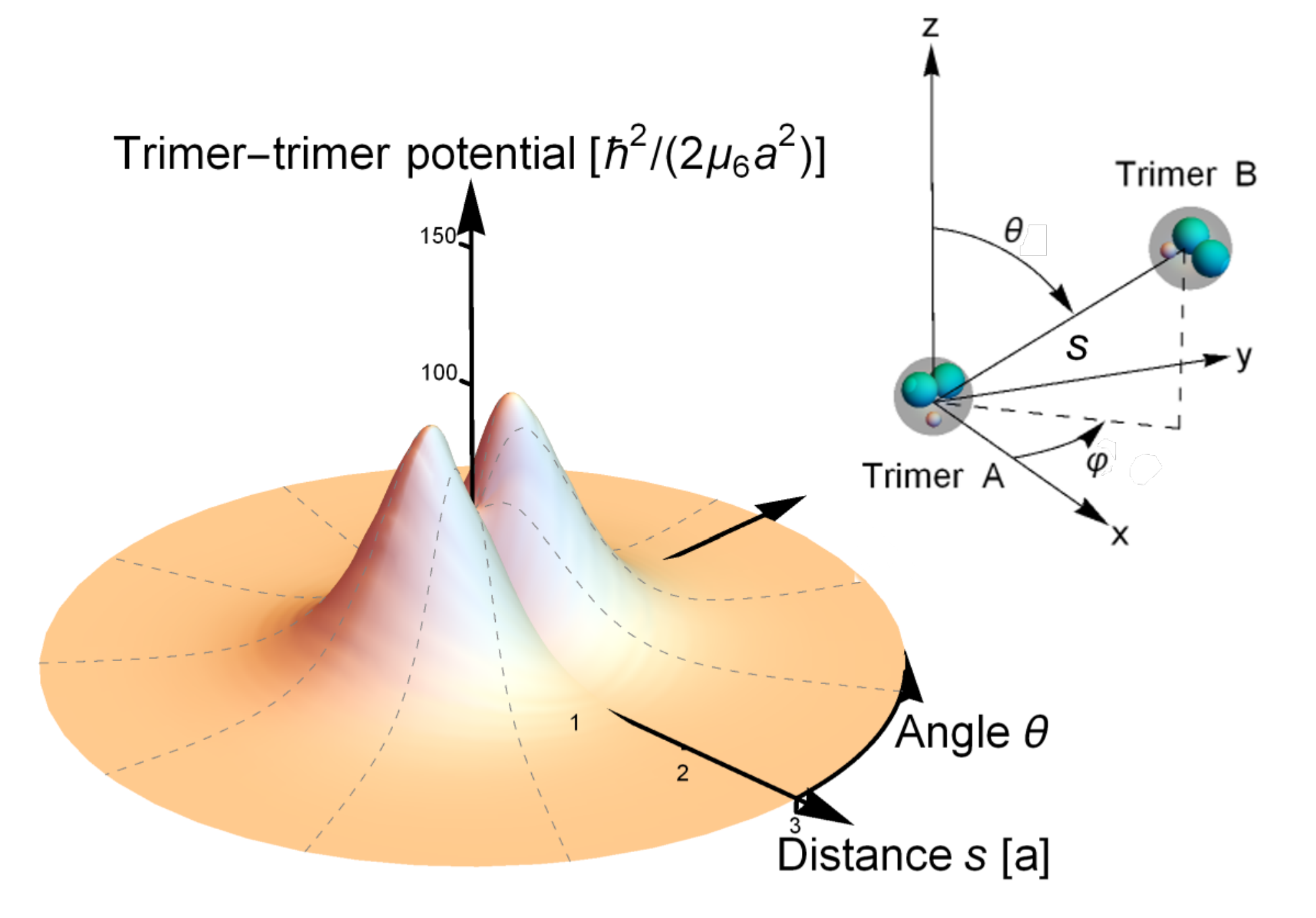}

\caption{\label{fig:Trimer-Trimer-potential}Integrated potential in the RGM1
- see Eq.~(\ref{eq:IntegratedPotential}) - between a trimer A in
rotational state $x$ and a trimer B in rotational state $y$, as
a function of the distance $s$ and angle $\theta$ (in the spherical
coordinates represented in the inset) between the centres of mass
of the two trimers.}
 
\end{figure}

\begin{figure}[b]
\includegraphics[scale=0.6]{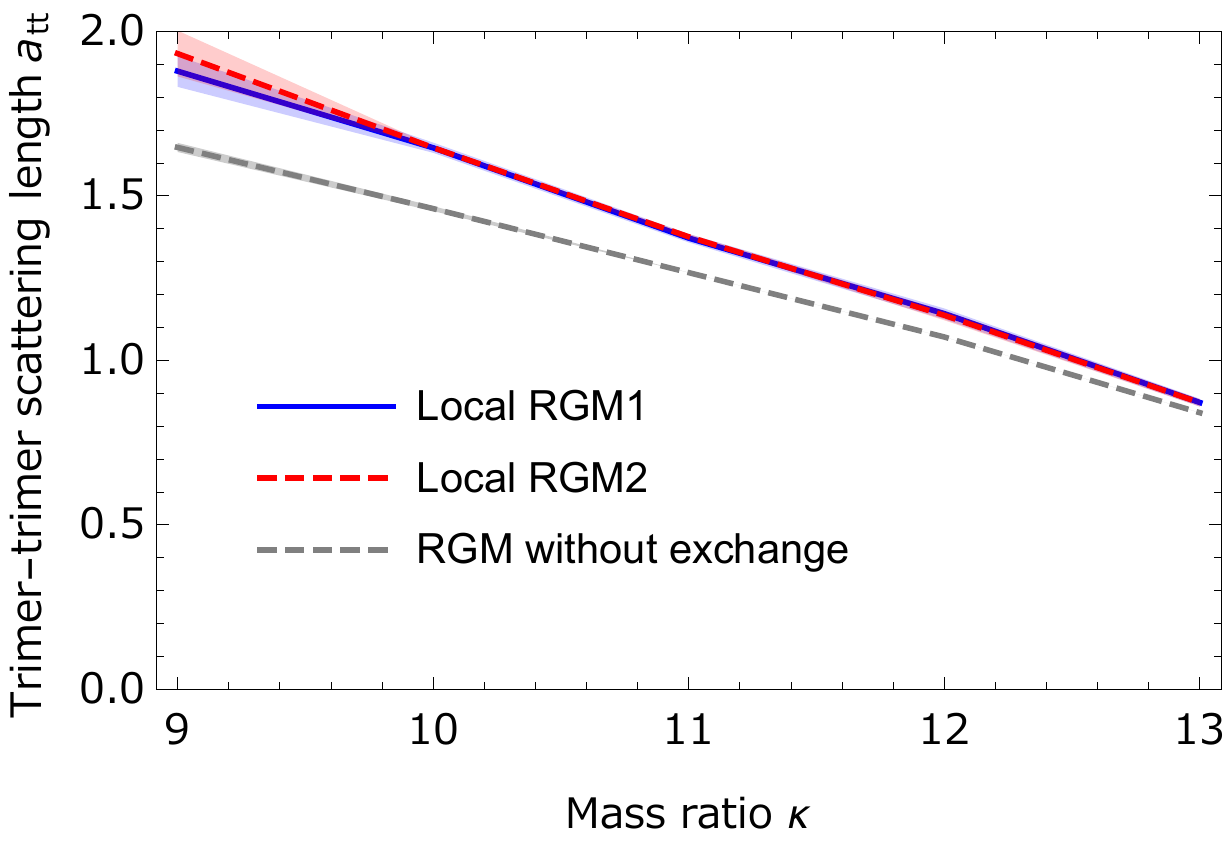}\caption{\label{fig:Trimer-trimer-scattering-length}Trimer-trimer scattering
length $a_{tt}$ in units of $a$ as a function of the mass ratio,
calculated by the local RGM1 and RGM2. The shaded area indicates the
statistical uncertainty due to the Monte Carlo integration used to
calculate the potentials.}
\end{figure}

The total potential (sum of direct and exchange potentials) 
\begin{equation}
V_{1}(\bm{s},\bm{s}^{\prime})=V_{D}(\bm{s})\delta^{3}(\bm{s}-\bm{s}^{\prime})+V_{\mbox{\tiny EX1}}(\bm{s},\bm{s}^{\prime})\label{eq:TotalPotential}
\end{equation}
 is anisotropic, due to the anisotropy of the trimers. To visualise
this anisotropy, we plot in Fig.~\ref{fig:Trimer-Trimer-potential}
the integrated potential
\begin{equation}
V_{\mbox{\tiny1 integrated}}(\bm{s})=V_{\mbox{\tiny D}}(\bm{s})+\int d^{3}\bm{s}^{\prime}V_{\mbox{\tiny EX1}}(\bm{s},\bm{s}^{\prime})\label{eq:IntegratedPotential}
\end{equation}
as a function of the distance $s$ and angle $\theta$ of the spherical
coordinates $(s,\theta,\varphi)$ - note that the potential does not
depend on $\varphi$ by rotational symmetry along the $z$ axis. Fig.~\ref{fig:Trimer-Trimer-potential}
shows that the anisotropy of the potential is moderate. As a result,
we only need to consider the partial waves $\ell=0$ and $\ell=2$
to get converged results. Fig.~\ref{fig:Trimer-Trimer-potential}
also indicates that the potential is repulsive. This fact is confirmed
by the numerical calculation of the potential in each partial wave
given by Eqs.~(\ref{eq:VDPartialWave}-\ref{eq:VEXPartialWave}).
As a result, the trimer-trimer $s$-wave scattering is positive.

We have found that the exchange potentials are to a good approximation
local potentials. In view of the previous results for dimers, we substitute
the exchange potential by their local approximation given by Eqs.~(\ref{eq:VEXPartialWave})
and (\ref{eq:localExchangePotential}) and neglect the exchange kernel,
which is costly to evaluate. We therefore use the local RGM1 and RGM2
equations, Eqs.~(\ref{eq:localRGM1}-\ref{eq:localRGM2}).

The resulting trimer-trimer $s$-wave scattering length is plotted
in Fig.~\ref{fig:Trimer-trimer-scattering-length} as a function
of the mass ratio $\kappa$. The results are similar to the dimer-dimer
case. As in the dimer-dimer case, the local RGM1 and RGM2 results
are very close, suggesting that the local approximation is enough
to reproduce the RGM, and the contribution from the exchange of particles
is small compared to the direct contribution. However, unlike the
dimer-dimer case, the scattering length decreases with the mass ratio.
This is due to the fact that the binding energy of the trimers increases,
and thus their size reduces, as the mass ratio increases. The decrease
of the scattering length is therefore a consequence of the decrease
of the scattering cross section, due to the decreased size of the
trimers.

\section{Conclusion}

We have applied the resonating group method to the scattering of universal
clusters which are described by zero-range interactions. We have found
that the single-channel RGM is relevant to clusters made of fermions.
It reproduces qualitatively, and in some limits quantitatively, the
exact results for scattering involving universal dimers. We have also
applied the single-channel RGM to the scattering of universal trimers.
It is found to be similar to the scattering of dimers: there is little
contribution from the exchange of particles and the effective interaction
is repulsive, unlike the scattering of a fermion and a dimer, where
exchange is dominant and produces attraction related to the Efimov
effect. As a consequence, we obtain a positive trimer-trimer $s$-wave
scattering length. This result has implications for the nature and
stability of the ground state of a mixture of heavy and light fermions
which are to be discussed in a separate work.

The validity and accuracy of the present RGM calculations are limited
by the single-channel approximation. In particular, it is likely that
trimers excite into the nearby dimer-particle continuum during their
collision, by analogy with nuclear systems where excited channels
play an important role~\cite{Hupin2013}. Including these extra channels,
i.e. states of the form Eq.~(\ref{eq:TotalWavefunction}) constructed
with other eigenstates of the $n$-body and $N-n$-body subsystems,
should converge to the exact results. It remains however numerically
challenging to go beyond the single-channel approximation for clusters
of more than two particles. As it stands, the single-channel RGM can
already give useful insights on the interactions between universal
clusters. It could be used to further investigate similar problems,
such as scattering of dimers and trimers, involving unpolarised fermions
or three-component fermions.

Acknowledgments: The authors are grateful to Tetsuo Hatsuda, Vojt\v{e}k
Krej\v{c}i\v{r}\'{i}k, Eite Tiesinga, and Sofia Quaglioni for helpful
discussions. P. N. acknowledges support from RIKEN through the Incentive
Research Project funding. S. E. acknowledges support from JSPS. A.
M. G. was supported by EPSRC, grant No. EP/I004637/1, FCT, grant PTDC/FIS/111348/2009
and a Marie Curie International Reintegration Grant PIRG07-GA-2010-268172.
\bibliographystyle{apsrev}
\bibliography{paperRGM}

\section{Appendix}

Here we give the explicit expressions for the variables appearing
in Eqs.~(\ref{eq:TrimerTrimerVEX1}) and (\ref{eq:TrimerTrimerVEX2}).

For Eq.~(\ref{eq:TrimerTrimerVEX1}) we have:
\begin{eqnarray*}
\bm{R}_{1} & = & \frac{(2\kappa+1)(2\kappa-1)}{2\kappa}\bm{s}-\frac{(2\kappa+1)^{2}}{2\kappa}\bm{s}^{\prime}+\frac{2\kappa+1}{\kappa+1}\bm{r}+\bm{R}\\
\bm{r}_{2} & = & (\kappa+\frac{1}{2})(\bm{s}^{\prime}-\bm{s})\\
\bm{R}_{2} & = & \frac{(1+2\kappa)(2\kappa^{2}-1)}{2\kappa(1+\kappa)}\bm{s}-\frac{(1+2\kappa)(2\kappa^{2}+2\kappa+1)}{2\kappa(1+\kappa)}\bm{s}^{\prime}\\
 &  & +\frac{1+2\kappa}{1+\kappa}\bm{r}+\bm{R}\\
\bm{r}_{3} & = & (\kappa+\frac{1}{2})(\bm{s}-\bm{s}^{\prime})+\bm{r}\\
\bm{R}_{3} & = & \frac{1+2\kappa}{2(1+\kappa)}(\bm{s}-\bm{s}^{\prime})+\bm{R}
\end{eqnarray*}
\begin{eqnarray*}
\bm{R}_{1}^{\prime} & = & \frac{2\kappa+1}{2\kappa^{2}}\bm{s}-\frac{(2\kappa+1)^{2}}{2\kappa^{2}}\bm{s}^{\prime}-\frac{2\kappa+1}{\kappa(\kappa+1)}\bm{r}+\bm{R}\\
\bm{r}_{2}^{\prime} & = & (1+\frac{1}{2\kappa})(\bm{s}-\bm{s}^{\prime})\\
\bm{R}_{2}^{\prime} & = & \frac{(1+\kappa-\kappa^{2})(1+2\kappa)}{2\kappa^{2}(1+\kappa)}\bm{s}-\frac{(1+2\kappa)(\kappa^{2}+3\kappa+1)}{2\kappa^{2}(1+\kappa)}\bm{s}^{\prime}\\
 &  & -\frac{1+2\kappa}{(1+\kappa)\kappa}\bm{r}+\bm{R}\\
\bm{r}_{3}^{\prime} & = & (1+\frac{1}{2\kappa})(\bm{s}^{\prime}-\bm{s})+\bm{r}\\
\bm{R}_{3}^{\prime} & = & \frac{1+2\kappa}{2(1+\kappa)}(\bm{s}-\bm{s}^{\prime})+\bm{R}\equiv\bm{R}_{3}
\end{eqnarray*}
\begin{eqnarray*}
\bm{r}_{1}^{\prime\prime} & = & -(1+\frac{1}{2\kappa})(\bm{s}-\bm{s}^{\prime})+\frac{1}{1+\kappa}\bm{r}-\bm{R}\\
\bm{R}_{1}^{\prime\prime} & = & -\frac{1+2\kappa}{2\kappa(1+\kappa)}\bm{s}-\frac{(1+2\kappa)^{2}}{2\kappa(1+\kappa)}\bm{s}^{\prime}\\
 &  & -\frac{1+2\kappa}{(1+\kappa)^{2}}\bm{r}-\frac{1+\kappa-\kappa^{2}}{\kappa(1+\kappa)}\bm{R}\\
\bm{r}_{2}^{\prime\prime} & = & \frac{1}{1+\kappa}\bm{r}-\bm{R}\\
\bm{R}_{2}^{\prime\prime} & = & -\left(1+\frac{1}{2\kappa}\right)(\bm{s}+\bm{s}^{\prime})\\
 &  & -\frac{1+2\kappa}{(1+\kappa)^{2}}\bm{r}-\frac{1+\kappa-\kappa^{2}}{\kappa(1+\kappa)}\bm{R}\\
\bm{r}_{3}^{\prime\prime} & = & \left(1+\frac{1}{2\kappa}\right)(\bm{s}^{\prime}-\bm{s})+\bm{r}\equiv\bm{r}_{3}^{\prime}\\
\bm{R}_{3}^{\prime\prime} & = & \frac{1+2\kappa}{2(1+\kappa)}(\bm{s}-\bm{s}^{\prime})+\bm{R}\equiv\bm{R}_{3}^{\prime}\equiv\bm{R}_{3}
\end{eqnarray*}

Note that $\bm{R}_{3}^{\prime\prime}=\bm{R}_{3}^{\prime}=\bm{R}_{3}$
and $\bm{r}_{3}^{\prime\prime}=\bm{r}_{3}^{\prime}$.

Additionally, for Eq.~(\ref{eq:TrimerTrimerVEX2}) we have:
\begin{eqnarray*}
\bm{R}_{4} & = & -\bm{R}-\frac{1+2\kappa}{1+\kappa}\bm{r}+(1+\frac{1}{2\kappa})(\bm{s}+\bm{s}^{\prime})\\
\bm{r}_{5} & = & -(\kappa+\frac{1}{2})(\bm{s}-\bm{s}^{\prime})\\
\bm{R}_{5} & = & -\frac{(2\kappa+1)}{\kappa+1}\bm{r}-\bm{R}+\frac{(2\kappa+1)^{2}}{2\kappa(\kappa+1)}\bm{s}^{\prime}+\frac{(2\kappa+1)}{2\kappa(\kappa+1)}\bm{s}
\end{eqnarray*}
\begin{eqnarray*}
\bm{R}_{4}^{\prime} & = & \bm{R}-\frac{1+2\kappa}{(1+\kappa)\kappa}\bm{r}-(1+\frac{1}{2\kappa})(\bm{s}+\bm{s}^{\prime})\\
\bm{r}_{5}^{\prime} & = & -(1+\frac{1}{2\kappa})(\bm{s}-\bm{s}^{\prime})\\
\bm{R}_{5}^{\prime} & = & -\frac{(2\kappa+1)}{\kappa(\kappa+1)}\bm{r}+\bm{R}-\frac{(2\kappa+1)^{2}}{2\kappa(\kappa+1)}\bm{s}^{\prime}-\frac{(2\kappa+1)}{2\kappa(\kappa+1)}\bm{s}
\end{eqnarray*}

And for Eq.~(\ref{eq:TrimerTrimerK}) we have:

\begin{eqnarray*}
\bm{r}_{4} & = & \bm{r}^{\prime}+(\kappa+\frac{1}{2})(\bm{s}-\bm{s}^{\prime})\\
\bm{r}_{4}^{\prime} & = & \bm{r}^{\prime}+(1+\frac{1}{2\kappa})(\bm{s}-\bm{s}^{\prime})
\end{eqnarray*}

\end{document}